\documentclass{sigchi}

\CopyrightYear{2018}
\setcopyright{acmlicensed}

\usepackage{balance}       
\usepackage{graphics}      
\usepackage[T1]{fontenc}   
\usepackage{txfonts}
\usepackage{mathptmx}
\usepackage[pdflang={en-US},pdftex]{hyperref}
\usepackage{color}
\usepackage{booktabs}
\usepackage{textcomp}
\usepackage{multirow}
\usepackage{float}
\usepackage{enumerate}

\usepackage{microtype}        
\usepackage{ccicons}          

\usepackage{todonotes}

\def\plaintitle{Effects of Automated Interventions \protect\\ in Programming Assignments: \protect\\ Evidence from a Field Experiment}

\def\emptyauthor{}
\def\plainkeywords{MOOC; intervention; exercise; programming}

\makeatletter
\def\url@leostyle{%
  \@ifundefined{selectfont}{
    \def\UrlFont{\sf}
  }{
    \def\UrlFont{\small\bf\ttfamily}
  }}
\makeatother
\def\pprw{8.5in}
\def\pprh{11in}

\definecolor{linkColor}{RGB}{6,125,233}
\hypersetup{%
  pdftitle={\plaintitle},
  pdfauthor={\emptyauthor},
  pdfkeywords={\plainkeywords},
  pdfdisplaydoctitle=true, 
  bookmarksnumbered,
  pdfstartview={FitH},
  colorlinks,
  citecolor=black,
  filecolor=black,
  linkcolor=black,
  urlcolor=linkColor,
  breaklinks=true,
  hypertexnames=false
}

\usepackage{tikz}
\newcommand\copyrighttext{%
  \footnotesize \textcopyright~Paper Authors 2018. This is the author's version of the work. It is posted here for your personal use. Not for redistribution. The definitive Version of Record was published in L@S 2018, http://dx.doi.org/10.1145/3231644.3231650. 
  }
\newcommand\authorcopyrightnotice{%
\begin{tikzpicture}[remember picture,overlay]
\node[anchor=south,yshift=10pt] at (current page.south) {\fbox{\parbox{\dimexpr\textwidth-\fboxsep-\fboxrule\relax}{\copyrighttext}}};
\end{tikzpicture}%
}

\begin{document}

\CopyrightYear{2018} 
\setcopyright{acmlicensed} 
\conferenceinfo{L@S 2018,}{June 26--28, 2018, London, United Kingdom}
\isbn{978-1-4503-5886-6/18/06}\acmPrice{\$15.00}
\doi{https://doi.org/10.1145/3231644.3231650}

\title{\plaintitle}

\numberofauthors{3}
\author{%
  \alignauthor{Ralf Teusner\\
    \affaddr{Hasso Plattner Institute}\\
    \affaddr{Potsdam, Germany}\\
    \email{ralf.teusner@hpi.de}}\\
  \alignauthor{Thomas Hille\\
    \affaddr{Hasso Plattner Institute}\\
    \affaddr{Potsdam, Germany}\\
    \email{thomas.hille@student.hpi.de}}\\
  \alignauthor{Thomas Staubitz\\
    \affaddr{Hasso Plattner Institute}\\
    \affaddr{Potsdam, Germany}\\
    \email{thomas.staubitz@hpi.de}}\\
}

\maketitle

\authorcopyrightnotice

\begin{abstract}
A typical problem in MOOCs is the missing opportunity for course conductors to individually support students in overcoming their problems and misconceptions.
This paper presents the results of automatically intervening on struggling students during programming exercises and offering peer feedback and tailored bonus exercises.
To improve learning success, we do not want to abolish instructionally desired trial and error but reduce extensive struggle and demotivation.
Therefore, we developed adaptive automatic just-in-time interventions to encourage students to ask for help if they require considerably more than average working time to solve an exercise.
Additionally, we offered students bonus exercises tailored for their individual weaknesses.
The approach was evaluated within a live course with over 5,000 active students via a survey and metrics gathered alongside.
Results show that we can increase the call outs for help by up to 66\% and lower the dwelling time until issuing action.
Learnings from the experiments can further be used to pinpoint course material to be improved and tailor content to be audience specific.
\end{abstract}





\section{Introduction}\label{ch:introduction}
Major differences when comparing MOOCs with in-class courses are the amount of students enrolled in MOOCs and the absence of direct personal communication.
Both differences make it difficult for instructors to support struggling students.
Individual feedback requires far too much time, not available to course conductors.
A suitable alternative might be peer assessments, but they usually require considerable effort to set up as well as a large amount of the students' time and can therefore only be applied once within a typical course runtime~\cite{Kulkarni2015}. 

Courses aiming to convey basic programming skills need a highly structured course corpus with hierarchical dependencies.
This is usually given with a carefully designed flow of units, starting at a very limited basis of given basics, such as being able to enter code into an editor in a structured manner, being encouraged to try things out and being capable of re-setting the environment to start anew.
From that on, all concepts to be conveyed have to be taught step by step, at best with constant exercises for application.
If a problem or misconception arises, appropriate feedback, if possible on individual basis, should be supplied to help students achieve their goals.

In-class education allows teachers to glance over the shoulder of their students to notice potential struggle, assess the current situation and give direct feedback.
Within online courses, this glance over the shoulder is not possible on large scale, therefore requiring an automated and scaling solution to detect struggling students and supply them with helpful feedback and additional training options.
In this study, we are testing three types of just-in-time interventions: (1) an encouragement to request feedback, (2) an encouragement to take a study break, and (3) the provision of bonus exercises.
With our experiments, we aim to answer the following research questions: 

\begin{enumerate}[$\hspace{20pt}$]
\itemsep-0.2em 
  \item [RQ1.] Does encouragement to request feedback or take a study break affect student behavior?
  \item [RQ2.] Do any of our just-in-time interventions affect course scores or dropout rates?
  \item [RQ3.] Do students value tailored or untailored bonus exercises?
\end{enumerate}


\section{Related work}\label{ch:relatedWork}
This paper contributes to the research area of dropout prevention and interventions in MOOCs.
We further recommend tailored exercises to students and are therefore researching in the area of recommender systems, student knowledge modeling and computer adaptive testing in the context of e-learning and MOOCs in specific.

\subsection{Dropout Prediction}\label{subsec:stopout}

Dropouts are students who stop participating in the MOOC they enrolled in.
Usually researchers take the last event, such as logging in, of a student as the date of a dropout~\cite{Taylor2014Likely,Yang2013Turn}.
MOOCs in general show the same phenomenon: roughly between 70 and 90 percent of all enrolled students do not finish the courses~\cite{Onah2014Dropout}.
Students take courses because of different reasons such as general interest, job relevance, the wish of earning a certificate or others~\cite{Kizilcec2015Motivation}.
In general it is difficult to find the exact reason why a student dropped out from a course, e.g. time constraints or lost interest, since the response rate of such surveys is generally low, between 12.5\%~\cite{KizilcecAttrition2015} and 1\%~\cite{Whitehill2015Beyond}. 
A general lack of time is the most common factor, as Kizilcec and Halawa stated in~\cite{KizilcecAttrition2015}, with a share of 84\% of participants mentioning that reason.
Other reasons could be that the students had problems in the practical exercises, or felt bored~\cite{Onah2014Dropout}.
In this paper we particularly aim at students who are interested in finishing the course but felt overwhelmed with the exercises.
In order to detect dropouts, supervised learning techniques like support vector machines, hidden markov models or logistic regressions have been used~\cite{He2015Identifying,Jiang2014Predicting,Kloft2014Predicting,Taylor2014Likely}. 
The exact features used to classify dropouts vary between the approaches but mostly consists of a mix of \textit{clickstream data}, \textit{grades}, \textit{social network analysis}~\cite{Yang2013Turn} and \textit{biographical information}.
While the classifiers are considerably accurate (85\% to 91\% within the same course~\cite{Whitehill2015Beyond}) the cited work often lacks suggestions on how to prevent students from dropping-out.

We want to address this issue by intervening on students who struggle with our exercises.
As suggested by Taylor et al.~\cite{Taylor2014Likely}, we focused on features which relate students to other students as such features are more predictive than average grades or login durations.
In accordance with Jiang et al.\cite{Jiang2014Predicting}, we aim to improve social integration and interaction by intervening struggling students with our so called social Request for Comments (RFC) (see Section~\ref{subsubsec:rfcInterventions}) feature.

\subsection{Interventions}\label{subsec:interventions}
The work of Carini et al.~\cite{Carini2006Student} and Yang et al.~\cite{Yang2014Question} supports the claim that course-related discussions of students can have a positive influence on student learning results.
As students can not be approached individually and directly, Chaturvedi et al.~\cite{Chaturvedi2014Predicting} and Chandrasekaran et al.~\cite{Chandrasekaran2015Learning} propose to use a machine learning approach to detect threads in forums which need to be addressed by the instructors.
 Agrawal et al.~\cite{Agrawal2015Youedu} also try to reduce the work for instructors by identifying confusion in forum threads and automatically suggesting a ranked list of suggested one-minute-resolution videos.
However, this type of intervention also requires students to ask questions in the forums.
Since students do not always reach out for help, being it because they do not realize that they need help~\cite{alevenLimitations2000} or because they are too shy, instructors can not help them directly in most cases.
Especially weaker students hesitate to reach out for help in forums according to Onah et al.~\cite{Onah2014Dropout}, who also mention that there is a lack of research on instructor-less, personal interventions within MOOCs. 
Therefore, researcher try to automatically detect students who are at risk of dropping out and intervene on them in order to motivate them.
Whitehill et al.~\cite{Whitehill2015Beyond} used their dropout predictions to survey stopped out students for feedback about why these students left the course.
They noticed that the surveys motivated students to come back sooner than those students who received no survey. 
Mailing specific subgroups of a course with targeted content may also increase the effect of an intervention~\cite{teusner2017InformedAction,Renz2016,kizilcec2014encouraging}.

These works show that small interventions, like sending emails to students, can have positive impact on student behavior.
Oftentimes, the effect of interventions might however also be non-existent, or at least not verifiable, as shown in a review by Kizilcec and Brooks~\cite{kizilcecDiverse2017}.
This is especially likely if the interventions are either not suitable or far too late to help students who struggle with exercises and thus face demotivation. 
When a rather generic intervention reaches the students days later, which does not help them with the actual problem they were facing, it is unlikely that they will come back.
Pre-emptive interventions, such as plan making interventions~\cite{Yeomans2017}, affirmation interventions~\cite{Kizilcec2017affirmation} and self-regulation interventions~\cite{Kizilcec2017PNAS} are suited to prevent undesired behavior before negative consequences come to effect.
While these interventions are administered at the start of a course, in this study we focus on just-in-time interventions in the context of programming assignments.
Specifically, students who are flagged as struggling while working on a programming problem receive an encouragement to either submit a Request for Comments (RFC) or take a study break.

\subsection{Recommender Systems}\label{subsec:rel:recommendation-systems}
Recommender systems are often used to improve the learning experience of students in MOOCs.
Bauman et al.~\cite{Bauman2014Recommending} identified knowledge gaps in student knowledge and proposed an algorithm to recommend remedial learning material they crawled from the Internet.
To predict the topics for which students have knowledge gaps they use the number of points a student received on a quiz and derive a score for each topic associated with the quiz.

Our approach shares many concepts with the aforementioned.
In order to determine deficits in content understanding, we also rely on annotation of learning material and quizzes (in our case, programming exercises).
In contrast to quizzes, programming exercises in introductory MOOCs are often pursued until a full score is achieved as they can be reattempted as many times as students wish. 
Therefore we added the metric how long students needed to complete the exercise compared to other students.

Recommender systems have also been used to recommend learning items such as websites, articles, books or exercises in~\cite{Michlik2010Exercises,Segal2014Edurank}.
Particular interesting for our work is the research from Michl\'{i}k et al.~\cite{Michlik2010Exercises}, as they dealt with recommending exercises to students in an e-learning context.
They annotated the learning objects with concepts and created a vector-space model to map the knowledge of the students.
After a student solved an exercise, she was asked to provide feedback which  is then used to calculate the knowledge model using computer adaptive testing (CAT).
From their work we adapted the idea of concept tagging and creating a vector-space knowledge model.
Since their feedback system relied on user feedback and not automated tests, we developed a new knowledge model suited for our specific use case and the increased user base within a MOOC.

EduRank developed by Segal et al.~\cite{Segal2014Edurank} used collaborative filtering in order to create a personalized difficulty ranking for exercises.
However, their algorithm works best if students have only partial overlap of accessed exercises.
This is not given in MOOCs because all students follow the same order of exercises, making it difficult to use in this domain.


\section{Study Design}\label{sec:concept}
This paper presents two types of interventions that aim to improve learning success, improve satisfaction and lower overstrain induced dropout of students of online programming courses.
The first type of interventions, just-in-time interventions in problem solving, motivates students to ask for help and feedback when they face an extended struggling period with an exercise.
The second type of interventions, bonus exercises, offers each student additional training exercises which are suited to tackle the individual weaknesses of the student.

\subsection{Just-In-Time Interventions}\label{subsec:interventions}
To prevent dropouts and increase student satisfaction, we want to help struggling students while they are working on their exercises. 
As shown in~\cite{Teusner2017Aspects}, students often struggled before they dropped out. 
From forum comments and other feedback we know that spending too much time is a cause of student dissatisfaction with a course. 
Therefore, helping students in the moment when they are stuck is beneficial, allowing to solve their problems before they lose interest and drop out. 

We issue just-in-time interventions to students when we think that they are struggling with an exercise. 
The purpose of these interventions is to interrupt the student and motivate them to rethink their approach, review video lectures, ask for help or do other exercises first.

\subsubsection{Identifying Struggling Students}\label{subsubsec:identifyStrugglingStudents}

\begin{figure}
\centering
    \includegraphics[width=0.3\textwidth] {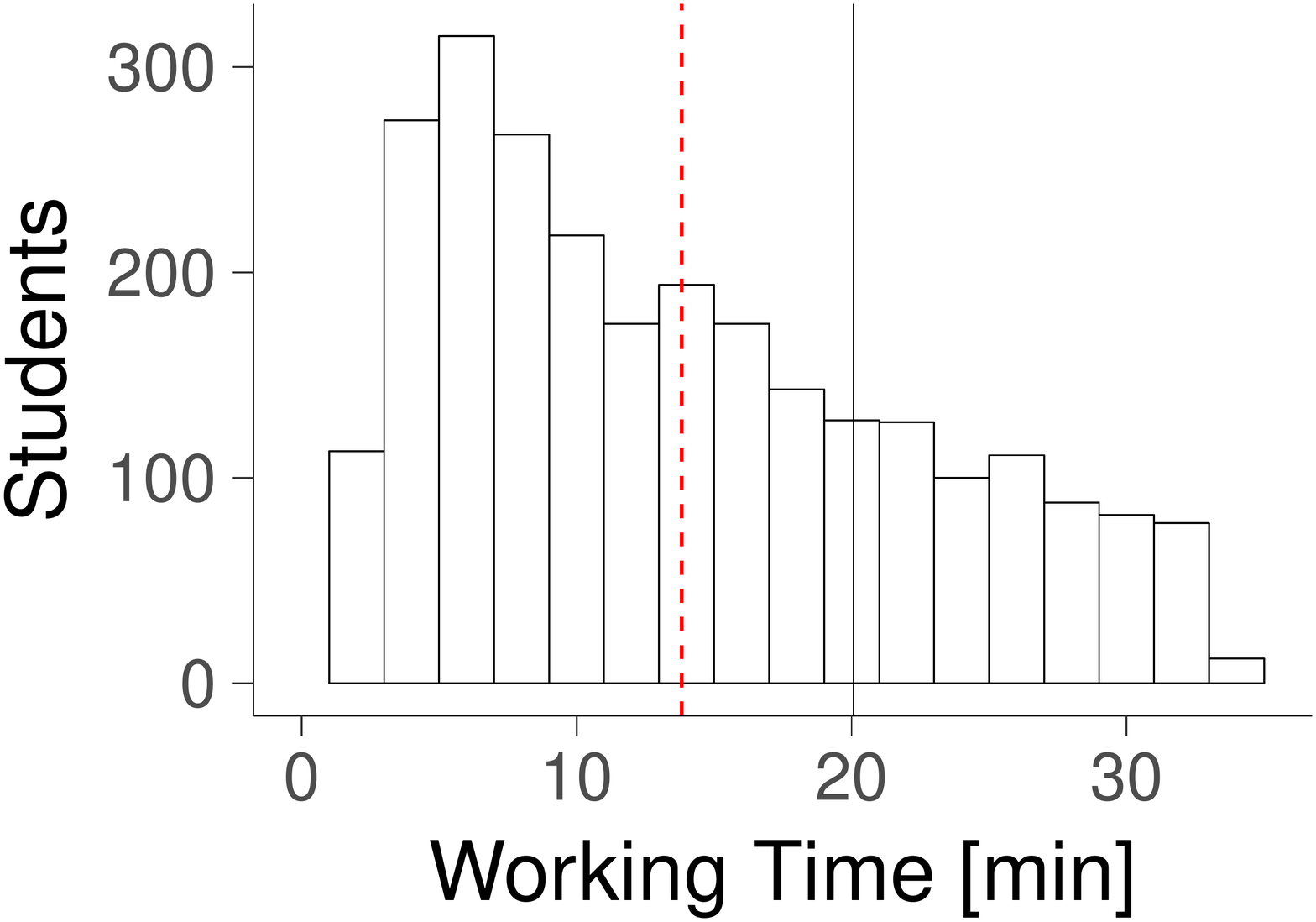}%
    \caption{Working time distributions of an exercise.}
    \label{pic:workingTimeSlow}
\end{figure}

To help struggling students we first need a method for identifying them.
The most obvious metrics for that are number of program executions and working times.
As the working times correlate strongly (Pearson coefficient of 0.9) with the amount of code executions and probably better reflect frustration levels, we decided on using working times.
We defined the 75th percentile of students to be regarded as considerable slower than their peers, therefore being potentially struggling or stuck and issued interventions on them.
This decision was backed by an analysis of the working times of already existing exercises from a prior course.

In Figure ~\ref{pic:workingTimeSlow} we see students' working times on a rather complicated exercise from the latest Java course with outliers removed.
The distribution is skewed to the left, meaning that most students finish the exercises quicker than the average duration (dotted vertical line), most likely without encountering any problems. 
The 75th percentile (solid vertical line) of working time was chosen as a cutoff to keep a relatively large treatment group in order to gather data, with the accepted risk of an increased ratio of false positives, interrupting students who are not yet struggling.
Other potential cutoff metrics which we considered but discarded were fixed times, averages, and lower percentiles (like the median).
Fixed times do not qualify as a suited cutoff metric, as the exercises are of different difficulty and complexity, thus the time needed to solve them varies and can hardly be predicted beforehand. 
Taking the average as a metric would classify too many students as struggling students and further has the problem that outliers have too significant impact and need to be removed.
The median (50th percentile) is not suited because we only want to indicate students who are significantly slower in solving the exercises.
Other metrics, such as taking a closer look on the code structure or detecting patterns on submissions, for example bursts on trying to run erroneous code as an indicator for frustration propose worthwhile future work.
This approach is thus not perfect, but allowed us to address the following problems:

\begin{enumerate}
\itemsep-0.1em
  \item \label{enum:FP} If exercises are short, we might bother students even though they are not struggling (false positives).
  \item \label{enum:coldstart} If only few students have worked on the exercise, data is still unreliable (cold start problem).
  \item \label{enum:lostfocus} If students review lecture videos while they work on an exercise, they should not receive interventions.
  \item \label{enum:continueExercises} If a student closes the exercise and comes back at a later point in time, when do we issue the intervention?
  \item \label{enum:annoyance} How to keep annoyance of interventions low and still cause an effect and gather data?
\end{enumerate}

To solve problems \ref{enum:FP} and  \ref{enum:coldstart}, we set a minimum time limit of 10 minutes.
If the 75th percentile lies below 10 minutes, we trigger the intervention at the 10 minute mark to let them solve their problems on their own first.

With regards to problem \ref{enum:lostfocus}, we just count working time while students are actively working on the exercise to prevent unnecessary or away from keyboard interventions. 
We stop the timer if the exercise view lost focus and continue if the student comes back to the exercise.

When students close the exercise and come back later, they need some time to understand the task and their previous code again.
For returning students, we therefore set the intervention interval to be the maximum of the remaining timespan to the 75th percentile and 10 minutes.

As we are aware that the interventions can annoy students, especially if they are not stuck or just like to fiddle with their code (\ref{enum:annoyance}), we set a daily limit of interventions to three interventions per day and student and an additional limit of two interventions per student and exercise.
This ensures that we intervene often enough to gather data and do not annoy the students too much.
Whether or not a student is actually struggling can not be detected automatically and will be covered by a survey.
The results shown in Section~\ref{sec:results} suggest that this rather simplistic approach appears to have been reasonably effective for our experiments.

\subsubsection{RFC Interventions}\label{subsubsec:rfcInterventions}
Requests for Comments (RFCs) are a useful feature for students to get help from other students who already solved the exercises.
A copy of the current status of an exercise is published internally, that enables other students to comment directly on specific lines of the code. 
It allows struggling students to reach out to their peers directly, which is more convenient than copying the code to the discussion forum. 
Unfortunately we have seen little use of this feature in previous courses. 
Although the feature was deemed helpful by the students using it, they seemed to need a little push to actually reach out for help, reinforcing the findings of Aleven and Koedinger~\cite{alevenLimitations2000}.
We do this by directly showing them the Request for Comments dialog. 
While encouraged by us, writing an RFC stays optional: the dialog is closable and might be reopened later, allowing the student to complete the present thought first.

\subsubsection{Break Interventions}\label{subsubsec:breakInterventions}
Break interventions encourage students to take a break and come back to the exercise with new ideas. 
The break interventions are also issued when we assume that the student is struggling with the exercise. 
Similarly, we show them a closable dialog, but here only presenting a text to remind them to do a break. 
Doing a break and giving the brain some rest can be beneficial to overcome side effects of concentrated working: fatigue and distraction, which results in errors and eventual frustration. 
Neuroscience researcher recommend taking a short break after periods of concentrated work~\cite{arigaBrief2011}.
Besides taking a break, we assume that struggling students also regard the break intervention as a motivation to review the course material.

\subsection{Bonus Exercises}\label{subsec:bonusExercises}
From previous courses we learned that many students want additional exercises for practicing. 
As some of our students told us that they had to leave the course because of time constraints, we cannot simply increase the amount of exercises for all students. 
Therefore we decided to add optional exercises to the course. 
Since we know how well a student performs in the course, we are able to recommend bonus exercises tailored to the strengths and weaknesses of the individual student accessing them.

Adding  bonus exercises as optional material has the benefit that students who want more exercises get more training and those who already spend enough time with the course are not penalized in terms of grading.

Learning material and exercises in MOOCs are often incrementally designed, which means that materials of ongoing weeks assume that students understood everything from the previous weeks.  
The problem might be that the student did not understand the exercise description. 
Thus, a differently designed exercise or repetition of the content might help them to understand the concept better. 

We posed a set of requirements for these exercises, of which the most important are:
the bonus exercises should be \textbf{optional}, meaning that students are not obligated to solve them and do not miss any graded points if they skip them.
They should be \textbf{solvable} for the student, meaning they should only cover topics already encountered.
They should be \textbf{tailored}, meaning they should deal with the concept the student had the greatest problem with.
They should be \textbf{focused} on the current course progress, so each course week will get its own pool of bonus exercises.
They should be \textbf{non-repetetive}, meaning that contrary to Michl\'{i}k et al.~\cite{Michlik2010Exercises}, we do not want to present an exercise to a student he already encountered another time, as we believe that the learning effect is low because students would either skip the exercise or copy the solution from the previous attempt.
And lastly, they should be \textbf{uncapped}, meaning that a student can request as many exercises as wanted.

Typically recommender systems present a list of most relevant items, i.e. exercises in our case.
We provide bonus exercises on a weekly basis. 
This means that each week has its own pool of potential bonus exercises in order to not confuse students with older material of previous weeks and to target specific deficits of the current week.
To find the most relevant bonus exercise, a content-based recommender system is used in favor of a collaboration based approach (see Section~\ref{ch:relatedWork}). 

As a prerequisite, we need to identify how well a student understood a topic. 
For this we manually annotated all exercises with the topics they cover and a difficulty level.
Having information about the student and his exercise submissions, we can create a vector-based profile, which is the basis of content-based filtering.

\subsubsection{Domain Model}
\begin{figure}
\begin{center}
    \includegraphics[width=0.8\columnwidth]{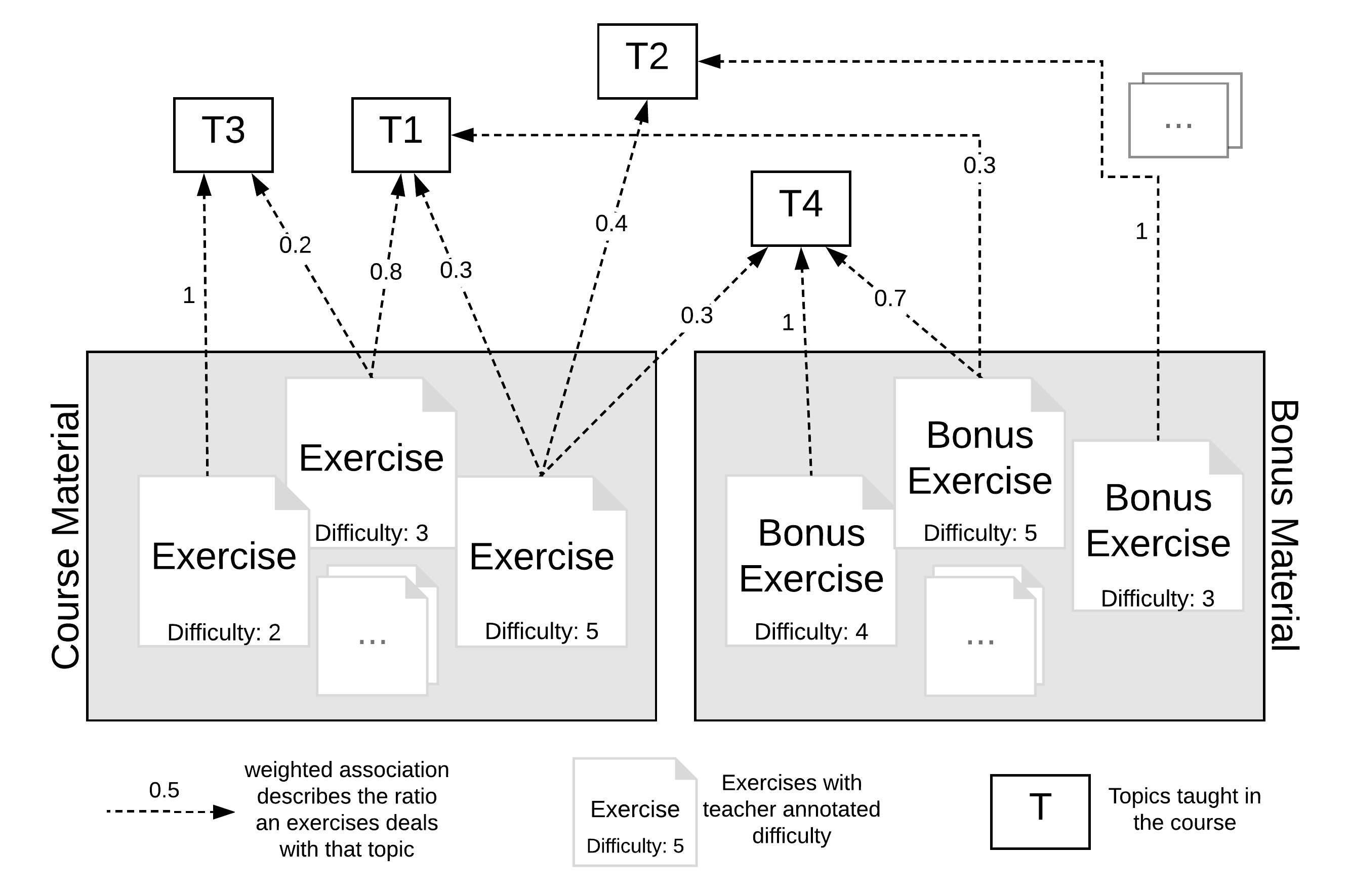}
    \caption{An example of our domain model.}
    \label{pic:ExerciseAnnotated}
\end{center}
\end{figure}

As a first step for our content-based approach, a domain model has to be defined.
It is important to note, that domain-models have to be created for each course independently. 
For example Java's class hierarchy is different from the one used in Python. To create the domain model, course instructors have to provide the topics students will learn in the course and annotate them to the exercises with which these topics are practiced. 
In Figure~\ref{pic:ExerciseAnnotated} we see that each exercise has one or more topics associated. We recommend not to use too many topics, i.e. less than three, and focus on the most important topics.
Weights are used to describe the importance of the topic in the exercise. Exercises are also annotated with their difficulty level.
Also we suggest to keep the topics as broad as possible and on the same abstraction level. 
In order to better represent exercises have a main topic but also cover other topics, associations (tags) may be weighted.
Michl\'{i}k et al.~\cite{Michlik2010Exercises} recommend to connect topics among themselves with directed relationships. 
To keep the additional effort for the teaching team small and not to integrate too many variables into our first experiment, we purposely chose to leave this task for future research.

\subsubsection{Knowledge Model}\label{subsubsec:knowledgeModel}
As a basis for recommendations, we create a profile in the form of a vector-space model for each student that reflects how well a student understood the topics taught in the course with a value between 0 (not understood) to 1 (fully understood). 
Similar to the model used in ALEF (Adaptive LEarning Framework)~\cite{Simko2010Alef,Michlik2010Exercises}, we are dealing with programming exercises.
However, we do not rely on self evaluation, but leverage our data on automated grading and testing. 
We assume that understanding of a concept and being able to apply it in exercises are correlated, therefore the understanding is directly correlated to the exercise performance.

Our knowledge model should reflect the following criteria:
\begin{enumerate}
\itemsep-0.1em
    \item Students, who needed longer with exercises than their peers, understood the concepts behind the exercise not as well as their peers.
    \item Not reaching full points means that a student had problems with the exercise. However, students, who got full points but needed very long to solve the exercise were able to show their ability to use the concept in the end. Therefore, the knowledge level of topics solved 100\% correctly should always be higher compared to fast but incompletely solved topics.
    \item Difficult exercises need deeper understanding of the topics. Consequently, difficult exercises have a higher potential learning effect and should be weighted stronger in the model.
    \item The results, in this case the exercise proposals, should be comprehensible.
\end{enumerate}

We developed a knowledge score function (see Equation~\ref{eq:lossFunctionWODimishing} and Table~\ref{table:knowledgeVariables}).
We calculate the knowledge score $\varTheta(s,t)$ for each student $s$ and each topic $t$ of the course.

\begin{equation}\label{eq:lossFunctionWODimishing}
  \Theta(s,t) = \frac{\sum_{e \in \mathcal{E}_s}\sigma(s,e) \cdot \delta(e) \cdot \rho(t, e) \cdot \varphi(e,\mathcal{E}_s)} {\sum_{e \in \mathcal{E}_s}\delta(e) \cdot \rho(t, e) \cdot \varphi(e,\mathcal{E}_s)}
\end{equation}

\begin{table}[H]
  \begin{center}
  \resizebox{.48 \textwidth}{!}{
    \begin{tabular}{|llll|}
    \hline 
    $s\in \mathcal{S} $ & Student $s$ in $\mathcal{S}=\{1, \dots , S\}$  & $\rho(t,e)$ & Ratio of topic $t$ in exercise $e$\tabularnewline
    $t \in \mathcal{T}$ & Topic $t$ in $\mathcal{T}=\{1, \dots , T\}$ & $\sigma(s, e)$ & Scoring of $e$ for student $s$ \tabularnewline
    $e \in \mathcal{E}$ & Exercise $e$ in $\mathcal{E}=\{1, \dots , E\}$ & $\delta(e)$ & Difficulty level of exercise $e$\tabularnewline
    $\mathcal{E}_s \subseteq \mathcal{E}$ & User accessed exercise $e \in \mathcal{E}_s$ & $\varphi(e,\mathcal{E}_s)$ & Diminishing function\tabularnewline
    $\iota (e,\mathcal{E}_s)$ & Returns position of $e$ in $\mathcal{E}_s $ & $\varTheta(s,t)$ & Knowledge score function\tabularnewline
    \hline 
    \end{tabular}
    }
  \end{center}
  \caption{Overview of Variables in Knowledge Model}
  \label{table:knowledgeVariables}
\end{table}

The formula consists of the following parts: 
\begin{description}
\itemsep-0.1em
 \item[\bf Scoring Function] The scoring function $\sigma(s, e)$ (see Table~\ref{tab:scoringFunction}) calculates how well we think the student $s$ solved the exercise $e \in \mathcal{E}_s$, based on the score and the working time of the student.
We take the test score the student reached in the exercise (row) and compare the student's working time to the working time of his peers (column). 
 The achieved test scores and working times are rounded down to the next lower (if possible) block, e.g. if a student reaches 95\% of the possible points, we round down to 80\%.
 If a student reaches full test score (100\% of points), he will never get a score below 0.7 regardless of his working time.
 We give a lower score (0.6 or less) if a student did not finish the exercise for 100\% to strongly separate the scores of solved exercises from unfinished ones.
 
 \begin{table}
  \begin{center}    
  \begin{tabular}{c|c|c|c|c|c|}
  \multicolumn{1}{c}{} & \multicolumn{1}{c}{} & \multicolumn{4}{c}{Working Time Percentile}\tabularnewline
  \cline{3-6} 
  \multicolumn{1}{c}{} &  & < 40\% & < 60\% &  < 80\% & $\geq$ 80\%\tabularnewline
  \cline{2-6}
  \multirow{5}{*}{\rotatebox[origin=c]{90}{Score}} & < 40\% & 0 & 0 & 0 & 0\tabularnewline
  \cline{2-6} 
   &  $\geq$ 40\% & 0.2 & 0.2 & 0.2 & 0.1\tabularnewline
  \cline{2-6} 
   &  $\geq$ 60\% & 0.5 & 0.4 & 0.4 & 0.3\tabularnewline
  \cline{2-6} 
   &  $\geq$ 80\% & 0.6 & 0.5 & 0.5 & 0.4\tabularnewline
  \cline{2-6} 
   & 100\% & 1 & 0.9 & 0.8 & 0.7\tabularnewline
  \cline{2-6} 
  \end{tabular}
  
  \caption{Values of the scoring function $\sigma(s,e)$.}
  \label{tab:scoringFunction}
 \end{center}
\end{table}

\item[\bf Weighting] We rank the scores $\sigma(s, e)$ based on the share of the topic $t$ on the exercise $e$ with $\rho(t,e)$ and on the difficulty of the exercise with $\delta(e)$.
\item[\bf Diminishing] Recent exercises better reflect the actual knowledge status of a student. Initial misunderstandings might have been clarified in the progress of the course exercises. In order to accommodate for this, we add the diminishing function (\ref{eq:diminishingFunction}):
  \begin{equation}
  \resizebox{.29 \textwidth}{!}{
    $\varphi(e,\mathcal{E}_s)=\frac{1}{1+e^{\frac{-3}{0.5\cdot|\mathcal{E}_s|}}\cdot(i-0.5\cdot|\mathcal{E}_s)|)}$
    }
    \label{eq:diminishingFunction}
  \end{equation}
$\varphi(e,\mathcal{E}_s)$ is an adapted sigmoid function that ranks exercises based on the order $\iota(e,\mathcal{E}_s)$ in which they have been solved.
Since we have different amounts of exercises for different topics, we adapt to the amount of exercises $|\mathcal{E}_s|$.

\item[\bf Averaging and Normalizing] To compute the final score, we calculate the average of all factors and normalize it between 0 and 1.
\end{description}

\subsubsection{Recommendation Algorithm}\label{subsec:recommendationAlgorithm}

Our recommendation algorithm first ensures that students are capable of solving the presented bonus exercise so they will not be overburdened right away.
From the pool of potential bonus exercises, we remove all exercises that are either too difficult for the student (\emph{difficulty appropriateness}) or contain topics the student has not used yet (\emph{concept appropriateness}).
The potential benefit for each bonus exercise is assessed by re-calculating all affected topic scores under the assumption, that the student fully solves the bonus exercise in optimal time. The sum of the resulting deltas of the topic scores is the potential benefit which we use to rank the bonus exercises and recommend them to the students.

For recommendation we serve only the highest ranked exercise to the students instead of providing them the ranked list.
Many students want to solve all offered exercises with 100\% score, providing them with a list of exercises thus may have negative effects.
Also since we want to support students with exercises dealing with their biggest weakness, giving them a list of exercises may not fulfill this requirement as they may chose the exercises they found the easiest.
If the list of ranked exercises remains empty, which happens if the student did not attempt any exercise yet or all potential exercises are too difficult for him at this point, we return the easiest exercise of the pool.
Since the knowledge model is updated for each exercise the students attempt, the system can be asked to recommend more exercises if required.


\section{Method} \label{subsec:evalSetup}
We conducted A/B tests for the just-in-time interventions and the bonus exercise intervention in a live programming MOOC with randomly assigned students.
For our experiment, we developed 15 distinct bonus exercises and additional 4 dummy exercises. 
Each of the four course weeks had a pool of 4 to 5 bonus exercises of average difficulty and one, much easier, dummy exercise assigned.
Additionally, we conducted two surveys, one at course start to assess a self-stated skill level together with some theoretical questions about OOP, and one at course end to learn about the perceived effects of our interventions.

The course-end survey contained questions on students' focus (I ... solely concentrated on exercises / did something in parallel (emails, chats) / sometimes concentrated, sometimes not), on students' impressions on intervention timing (The interventions appeared... much too early / when I was stuck / after I solved the exercise / never), intervention helpfulness (The interventions... helped me / bugged me / did not bother me / never appeared) and satisfaction with the bonus exercises (The practical bonus exercises were... helpful and fitted my weaknesses / were good but not helpful / were too difficult / superfluous and I did not solve them).

\subsection{Participants} \label{subsec:evalParticipants}
5,839 students accessed at least one practical exercise, 1,663 (28.5\%) of them finished the course. 
The course-start survey was answered by 6,486 students, the course-end survey was completed by 1,257 students.
20.2\% of the students stated in the course-start survey that they never programmed before taking our course, 
32\% stated to have already rather good knowledge in programming, the remaining 47.8\% students had some general knowledge.
However, in the ungraded multiple choice test, 50\% to 80\% of the given answers were wrong.

\subsection{Measures} \label{subsec:evalMeasures} 
We measure the dropout rates, average scores and average working times of the students within the experiment and control groups.
For our concept of interventions we need to measure how long students need to solve an exercise.
Our coding platform saves events when students run their code (\emph{run}), run the unit tests (\emph{assess}), or submit their achieved score to the MOOC platform (\emph{submit}).
From these timespans, we can calculate the working time by summing up the timespans in between.
Students however might also need several attempts in different sessions, or  take breaks during their work.
Since additional auto-saves are done automatically after code changes, we can further cleanse the working time calculation by removing longer periods without changes, which we classify to be breaks.
Timespans longer than five minutes are discarded for the calculation, we expect the student having logged off in between, left the computer, browsing the web or other long term activities.
We argue that a focus loss of less than five minutes is still within the working time of the exercise, because a student may think about a problem without changing code.
Another important factor to exclude is that some students continue working on finished exercises to improve the style or try out different implementations.

Summarized, for the calculation of the working time we aggregate all time gaps of events which are less than five minutes and, if applicable, until the student first reached max score.

\subsection{Procedures} \label{subsec:evalProcedures}
To assess prior knowledge, we collected self-stated skill levels and conducted a short, ungraded multiple choice test about OOP and programming concepts.

To test the just-in-time interventions, we split the students into three disjunct groups: The control group (no interventions, 20\%), and two experimental groups (break, 20\% and RFC interventions, 60\%).
We have chosen to make the RFC group larger to generate more user data and more RFCs.
When measuring the effects of just-in-time interventions, we assume that RFCs sent within less than 10 minutes after the intervention was issued were affected by it.

To test the bonus exercise intervention, we split the same students into three additional groups: The control group (dummy exercises, 20\%) who received a bonus exercise which was mainly a copy of the first exercise of each week. 
This was necessary in order to prevent confusion in the forums why some students received bonus exercises and some did not.
The first experimental group (random exercises, 20\%) received random bonus exercises from our pool of bonus exercises.
This gives a baseline on how well our recommendation algorithm works.
The second experimental group (tailored exercises, 60\%) received tailored bonus exercises from our pool of bonus exercises.
In contrast to the random group, they are based on the knowledge model of the student.
The tailored bonus exercise group is again larger in order to generate more data.

We further surveyed the students at the end of the course on their perception of helpfulness and timing of the interventions as well as their focus on the exercises.

\subsection{Implementation of Interventions} \label{subsec:implInterventions}
As described before, we issue interventions after a calculated amount of time at which we assume that students struggle.
In order to keep the user interface as responsive as possible, we retrieve the needed information, the 75th percentile and working time of the student on the requested exercises, asynchronically in the frontend using Javascript.
Once the necessary data is collected by the frontend, a timer is started.
If the timer reaches zero, an intervention is shown.
As we do not want interventions to show up when the student is not really working on the exercise.
Therefore, the timer is automatically stopped if the browser focus of our coding platform  is lost, e.g. if the student is distracted to check emails or to perform other actions.
No interventions are shown after the student solved the exercise completely.

\section{Results} \label{sec:results}

\begin{figure}
\begin{center}
    \includegraphics[width=0.8\columnwidth]{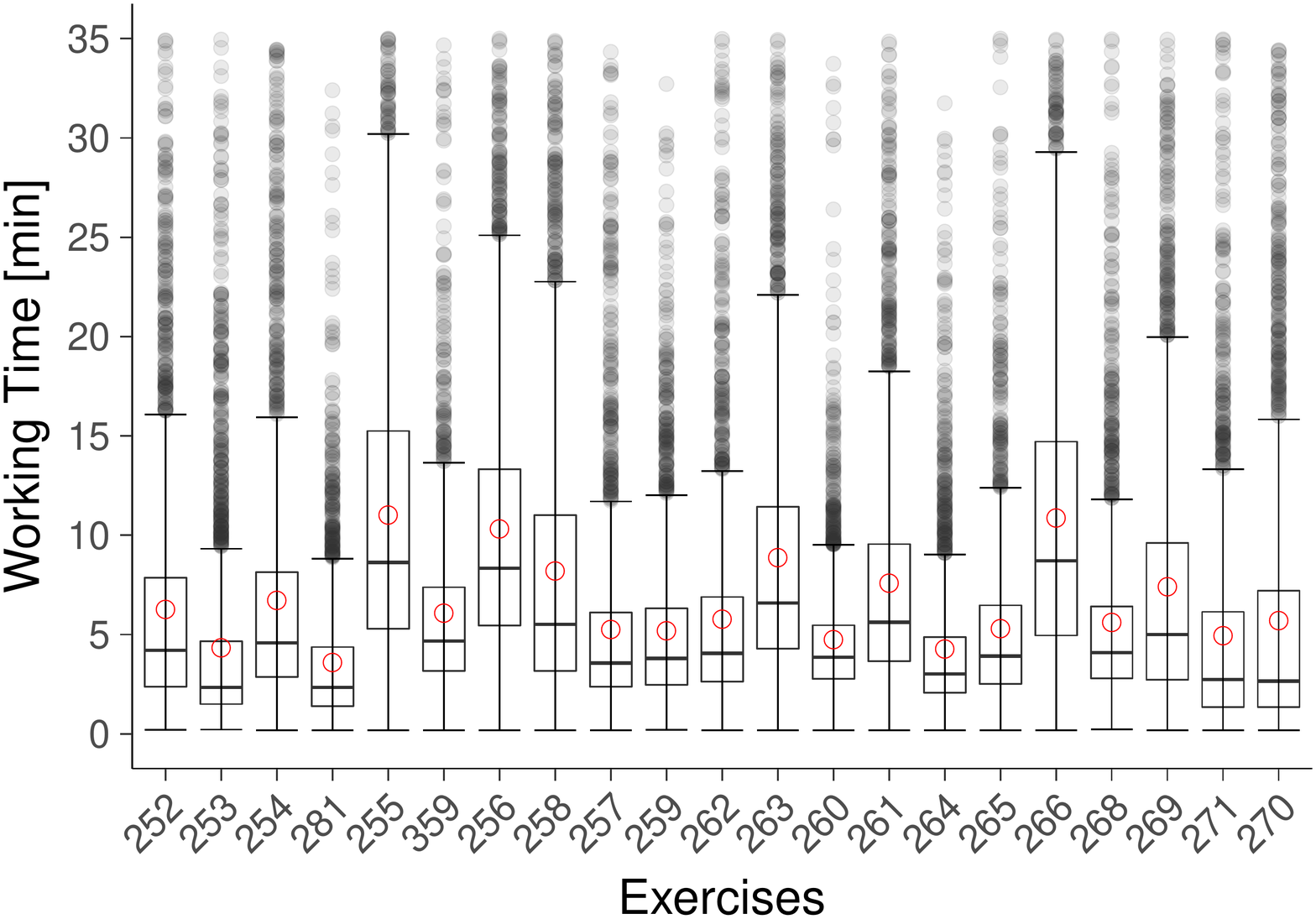}
    \caption{Working times of students (2nd week). Circles mark averages, black bars mark medians.}
    \label{pic:exerciseBoxplotsWeek2}
\end{center}
\end{figure}

\subsection{Descriptive Statistics} \label{subsec:descStatistics}
The average scores of the 65 course exercises range between 95\% and 99\%.
 This reflects the high diligence of the students to finish the exercises with full score.
As shown in Figure~\ref{pic:exerciseBoxplotsWeek2}, most of the students (75\%, indicated by the upper end of the boxes) had no problems solving the exercises (exemplary taken from week 2) in less than 15 minutes.
The slower working time distributions are skewed to the right (long whiskers on top), which we want to reduce with the help of just-in-time interventions.

\subsection{Just-in-Time Interventions} \label{subsec:evalInterventions}
During the experiment we sent 8,205 break interventions and 25,426 RFC interventions to the students.
79\% of all answered RFCs helped students to achieve full score, on the other 21\% the student either did not come back to the exercise or solved the problem before an answer was posted.
Table~\ref{table:keyFiguresInterventionsGroup} shows course dropout rates and scores in each experimental condition.
As the first of four key findings on the just-in-time interventions, it is apparent that they had no significant effects on dropout rates or scores in the course (all $p>0.69$).

\begin{table}
\caption{Course Key Metrics. Percentages are averages.}
\centering
\resizebox{\columnwidth}{!}{
    \begin{tabular}{lccccc}
    \hline 
    Group & \#started & \#finished & Dropout rate & Score all students ($\sigma$)  & Score finisher ($\sigma$) \tabularnewline
    \hline 
    \hline 
    No interventions & 1166 & 343 & 70,6\% & 46\% ($\pm$35\%) & 94\% ($\pm$10\%)\tabularnewline
    Break interventions & 1155 & 331 & 71,3\% & 44\% ($\pm$34\%) & 93\% ($\pm$12\%)\tabularnewline
    RFC interventions & 3518 & 989 & 71,9\% & 45\% ($\pm$34\%)  & 93\% ($\pm$14\%)\tabularnewline
    \hline 
    \end{tabular}
}
\label{table:keyFiguresInterventionsGroup}
\end{table}

Metrics for the just-in-time interventions are shown in Table~\ref{table:interventionsEffectiveness}.
In the control condition, 24\% (172/720) RFCs were sent in the time frame where an intervention would have appeared.
The break intervention had no effect on RFC timings, with 24\% (229/944) RFCs sent shortly after a break message. 
However, in the RFC encouragement condition, 31\% (1065/3448) RFCs were sent following the intervention.
Thus, as a second finding, the RFC intervention significantly increased the rate of RFCs by 7 percentage points relative to the control condition (Chi-squared test $\chi^2=25.32$, $p<0.001$). 

\begin{table}
\caption{RFCs and Break metrics after interventions. Percentages are averages.}
\centering
\resizebox{1\columnwidth}{!}{    
    \begin{tabular}{llcll}
        \hline 
        Group & RFCs/student & RFCs after intervention & Time to RFC & Break duration\tabularnewline
        \hline 
        \hline 
        No interventions (baseline) & 0.6 & 24\% & 32.1 {[}min{]} & 14.46 {[}min{]} \tabularnewline
        Break interventions & 0.8 (+33\%)& 24\% & 28.8 (-10\%)& 18.07 (+25\%)\tabularnewline
        RFC interventions & 1.0 (+66\%)& 31\% & 30.4 (-5\%)& 17.50 (+21\%)\tabularnewline 
        \hline 
    \end{tabular}
}
\label{table:interventionsEffectiveness}
\end{table}

\begin{figure}
\begin{center}
    \includegraphics[width=0.9\columnwidth]{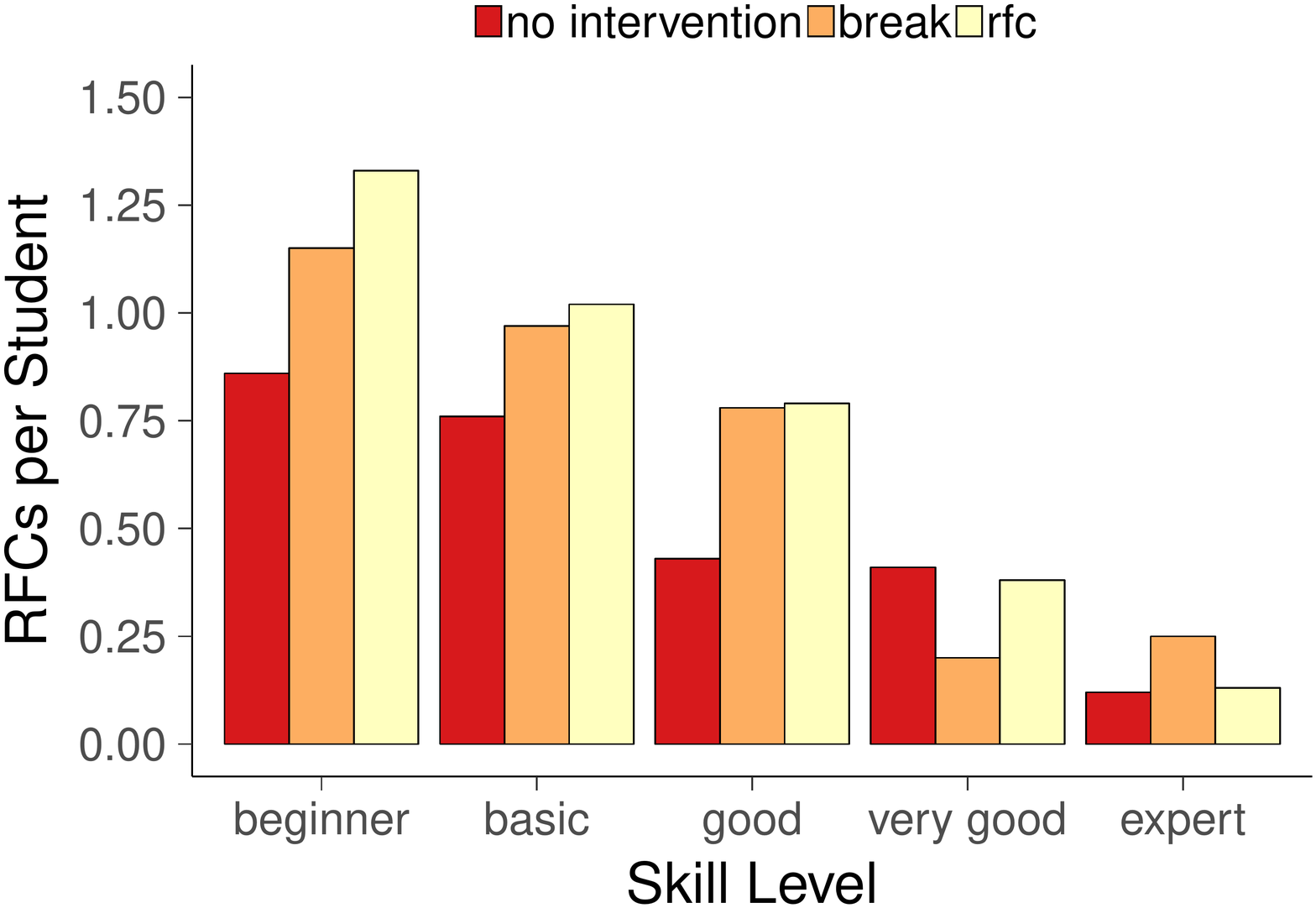}
    \caption{RFCs per Student and Skill Level}
    \label{pic:rfcPerSkillLevel}
\end{center}
\end{figure}\vspace{-5pt}

Taking into account students' self-reported skill levels, we further find that the effects of just-in-time interventions vary significantly across different skill groups, as can be seen in Figure~\ref{pic:rfcPerSkillLevel}.
The RFC and break interventions raised RFCs among less skilled students ('beginner' to 'good'), while the effects were mixed among highly skilled students.

\begin{figure}[h]\vspace{-5pt}
\begin{center}
    \includegraphics[width=0.66\columnwidth]{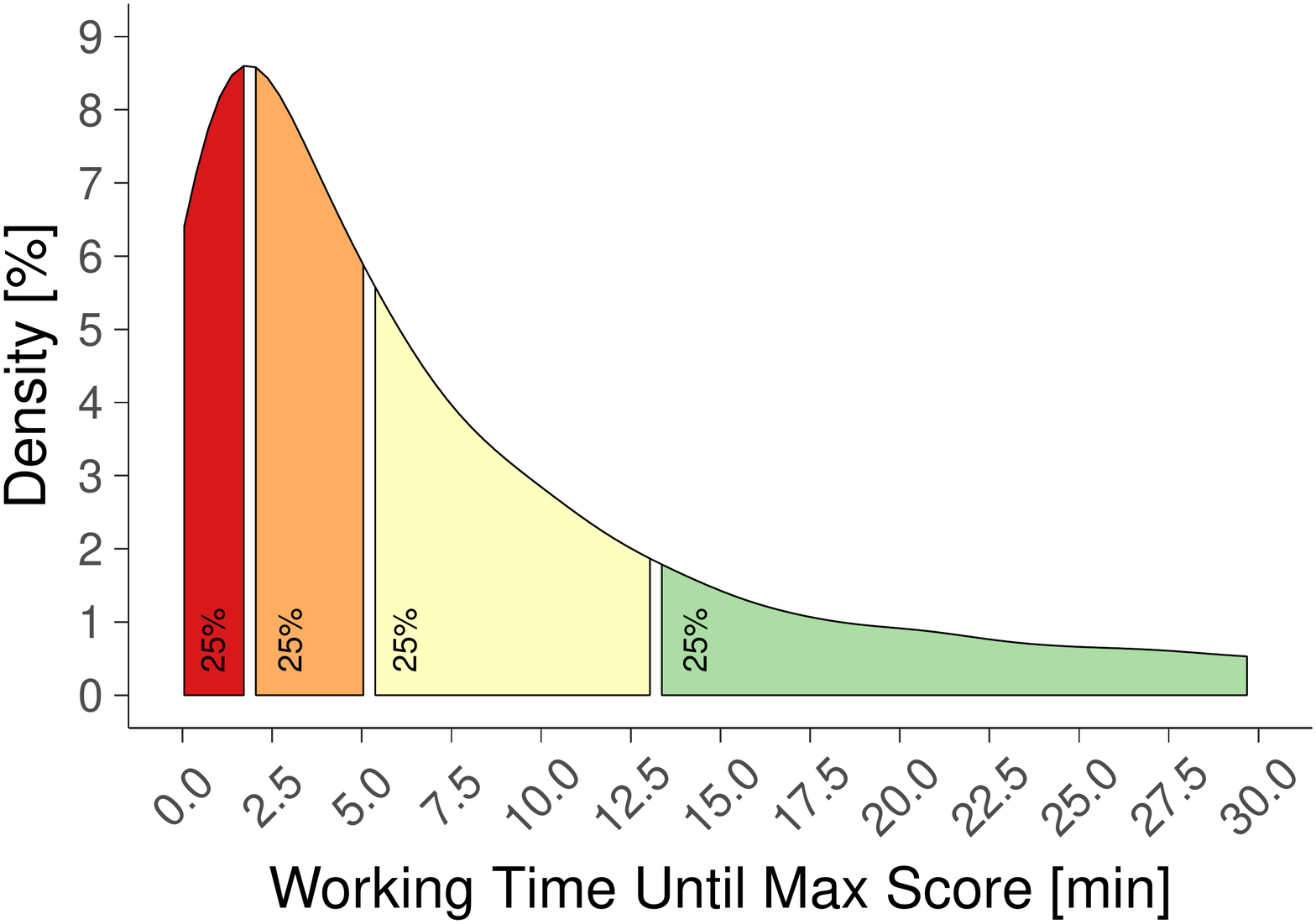}
    \caption{Density Plot: working time students need to reach full score after receiving an RFC reply. Each section represents 25\% of students.}
    \label{pic:density}
\end{center}
\end{figure}

The required working time to solve the exercise after a student received feedback is visualized in Figure~\ref{pic:density}.
Half the students finished their exercise within 5.2 minutes additional working time after they got a reply to their RFC. 
The data is further highly skewed to the left, which means that most students finish rather quickly, 25\% of the students could even resolve their problem in less than 2 minutes.

According to the course-end survey most interventions were not perceived as very helpful to the students, as a large share (47\%) answered they simply ignored them or felt bugged by them (30\%).
Considering the timing, 19\% answered that the timing was on spot, 53\% perceived interventions as too early, 7\% as too late or after solving the exercise, and stated to have never seen a just-in-time intervention. 
With regards to focus, most students stated that they worked concentrated on the exercises (70\%), 23\% were sometimes distracted, only 7\% were often distracted.

\subsection{Bonus Exercises} \label{subsec:resultsBonusExercises}
\begin{figure}[h]\vspace{-8pt}
\begin{center}
    \includegraphics[width=0.72\columnwidth]{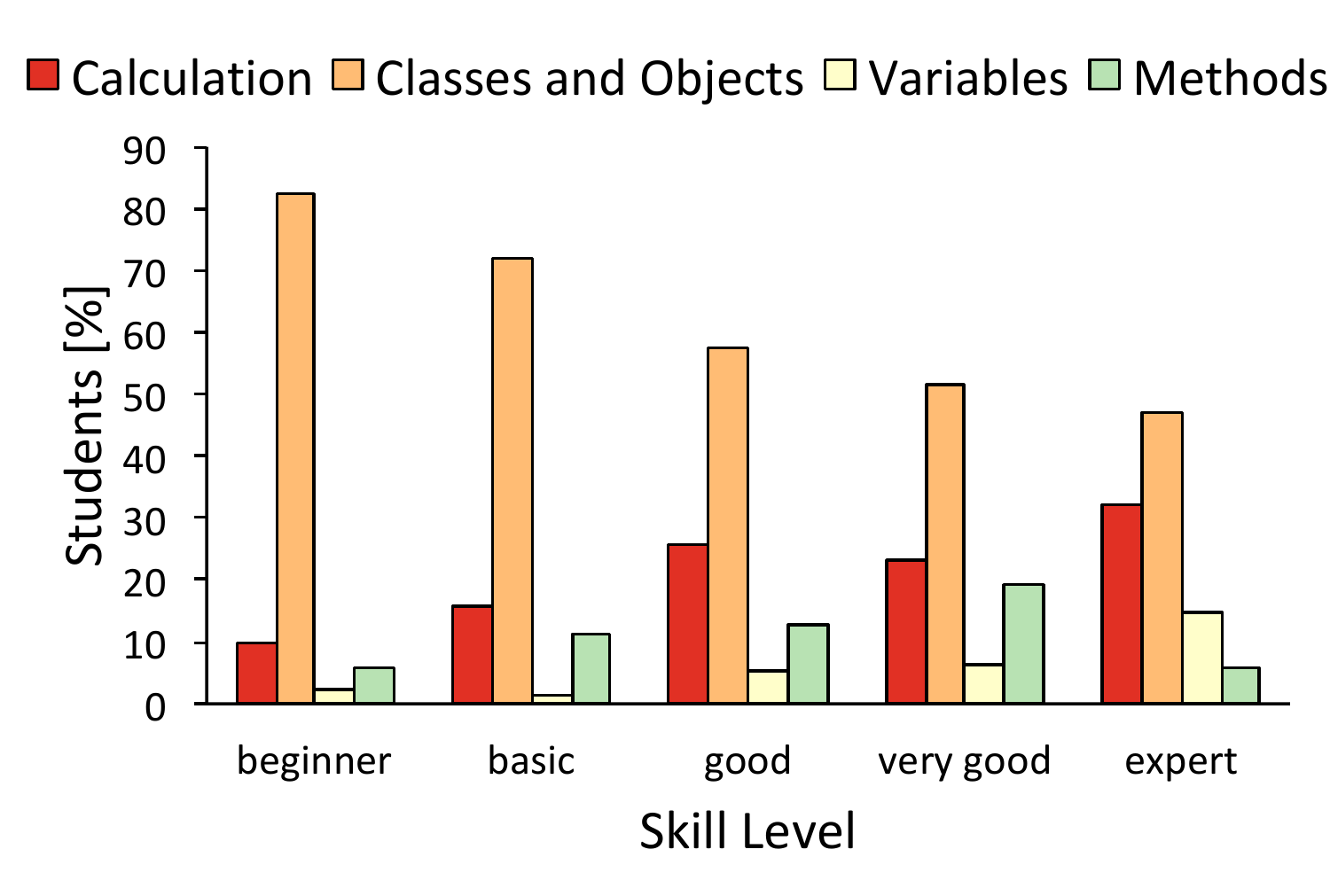}
    \caption{Students Weakest Topic of Week 1 by Skill Level.}
    \label{pic:weaknessesBySkillWeek1}
\end{center}\vspace{-8pt}
\end{figure}

Results show five key findings with regards to bonus exercises.
First, students interacted with the optional bonus exercises at a similar rate as with the standard exercises with regard to the starting rate (Welch Two Sample t-test $t=1.8$, $p=0.13$), completion rate ($t=-0.06$, $p=0.96$) and working time ($t=0.98$, $p=0.33$).
Second, when analyzing the differences in weaknesses specific to skill groups, certain topics seem to be harder than others for students stating low skill levels. 
With increasing skill levels, the distribution of weaknesses becomes more balanced (Figure~\ref{pic:weaknessesBySkillWeek1} shows that exemplary for course week 1).
Third, when comparing the working times of the untailored random group and the tailored group on longer bonus exercises, the tailored group needed 10.6 minutes on average, while the untailored group needed 11.7 minutes (10\% increase, we conducted a Wilcoxon signed-rank test, $p=0.39$, no statistical significance).
Fourth, student perceptions did not vary significantly based on whether the bonus exercises were tailored or untailored ($\chi^2=0.74$, $p=0.86$), however numbers differed for the control group with dummy exercises (all $\chi^2$ > 22, all $p<0.001$).
Fifth, going into detail, the bonus exercises were mostly received as helpful and fitting the specific weaknesses (59\%, result for control (dummy group): 39\%).
About 31\% (control~51\%) perceived the bonus exercises as good but not specifically helpful, 4\% (control~3\%) as too difficult, and 5\% (control~7\%) as superfluous in general.
The number of bonus exercises, one per week, was regarded as just right by the majority of students, about 21\% wanted more, less than 8\% fewer exercises.

\section{Discussion} \label{sec:discussion}
To evaluate our two concepts to support students in MOOCs, just-in-time interventions and recommended bonus exercises, we conducted A/B tests on a live course with 5,839 active students.
In this study, we found that just-in-time interventions have an effect on student behaviour (RQ1); the interventions had no measurable effect on course scores or dropout rates (RQ2); and students valuate bonus exercises (RQ3).

\subsection{Just-in-Time Interventions} \label{subsec:discInterventions}
Despite students stating in the survey on the helpfulness of RFCs and suggestions of their peers, there is no significant effect to scores or dropout rates (see Table~\ref{table:keyFiguresInterventionsGroup}).
A possible explanation could be that the proportion of students, who struggled and profited from the interventions, is too low within the group to significantly change the overall number.
Additionally, with the average reached scores of the exercises ranging between 95\% and 99\%, we likely encounter a ceiling effect, limiting the effective room for improvement. 
It might be worthwhile to repeat the experiment with a smaller share of students exposed to the interventions, for example only approaching the slowest 10\%, to reduce false positives and probably streamline our actually affected user group.

For all results from our survey, we need to consider that the survey was conducted at course end, therefore mainly targeting students who had fewer problems with the exercises and therefore did not benefit as much as weaker students from the interventions.
The relatively high share of answers that students felt bugged (30\%) might be caused by the fact that we had no limitation on interventions per day in the first half of the course.
With regard to the timing, we hit our goal of issuing too many interventions if in doubt.
A relatively large share (19\%) answered that the timing was on spot, albeit an even larger share answered that it was too early (53\%). 
These results also hint to increase the percentile after which to issue interventions and thus to affect fewer students.

The duration of breaks increases and students take action earlier, regardless of the type of the just-in-time intervention. 
This seems reasonable as both just-in-time interventions disrupt the student in the current task and might trigger a step back in mind.
Additionally, writing an RFC takes some time and thus prolongs the duration until next runs are issued.

We further see that, despite the fact that most students reported that they were unaffected by the just-in-time interventions, the experimental groups show an increased use of RFCs of up to 66\% in the RFC group (see Table~\ref{table:interventionsEffectiveness}, RFC per student).
Interesting to note is, that students who received a break notification showed a 33\% increase in their relative RFC amount as well.
A likely explanation is that students used the reminder to take a break as a hint to reach out for help in a form of RFCs.

In general, students send fewer RFCs with increasing prior knowledge (see Figure~\ref{pic:rfcPerSkillLevel}).
Students that already possess very good or expert knowledge did not further benefit from the just-in-time interventions.
The biggest effect was noticed on the beginner group, which hit our intentions and expectations.

\subsection{Bonus Exercises} \label{subsec:discBonusExercises}
Only 10\% of our students deemed the bonus exercises unnecessary, regardless of the experiment group they were part of.
Bonus exercises, even if repetitive, are therefore valued.
We do not see a difference in valuation between randomly assigned and tailored bonus exercises.
Even the dummy exercises were regarded as helpful, although to a smaller fraction than the purposeful exercises~(39\% vs. 59\%).
A possible explanation for this is that every exercise is considered helpful in a beginners course as it allows to train basic syntax.
A rather high share of 31\% answers stating that the exercises were ``good but not specifically helpful'' might be caused by the fact that the responders of the survey succeeded the course and therefore had or acquired some skill.

When comparing the working times of students of the random group and students of the tailored group on the same bonus exercise, we expected the tailored students to be slower, since they have to work on their weaknesses.
The opposite turned out to be true.
Probably, the cause of marking the topic as their weakness, a longer struggle and more tinkering, resulted in a better and deeper understanding for them.
However, since we could not conclude statistical significance we propose to further investigate on that finding.

The analysis of weaknesses per skill level reassures us that the algorithm in general is working as intended.
Exemplary, within the first week, the biggest issue seems to have been Classes and Objects, further hinting that either the exercises on that topic were too difficult or the explanations could be improved.
This finding is also backed by exercise metrics, having the lowest average score of that week and the highest number of RFCs.
The insights gained from the analysis thus can further be used to detect points for improvement and shortcomings in the learning material.


\section{Limitations and Future Work}\label{ch:future-work}
The results of our experiments have to be validated and are inherently limited by our experiment setup. 
In order to improve general conclusions, we have to identify and mitigate the influence of the instructor set variables, such as the chosen percentile to intervene on, topic weightings and difficulty ratings. 
For the just-in-time interventions, especially the selection of the intervention group, which relies on detecting struggling students, is likely to contain further potential.
Our current approach depending on percentiles can not separate actually struggling students from ones just taking it slow. 

In order to improve that and further individualize the feedback, we could analyze common programming errors and create interventions or hints specifically to these errors.
If students often run into the same error, we could try to help them with interventions tailored to the type of error they get, for example by pointing  to a specific video.
There is lots of research in the field of code analysis which would help us to identify struggling students and provide them with meaningful feedback~\cite{Jadud2006Methods}.
Students who show many typing errors (typos) might need to recap syntax and we could automatically display cheat sheets to them. 
One could also detect if students often run the same code with little to no change (bursts) and intervene them in order to prevent them from building up frustration.

Many students replied that interventions appeared too early, therefore we plan to make interventions smarter by recognizing progress.
We will further try to predict when students usually do RFCs based on metrics such as errors or size of code changes to improve intervention timing.

The impact of the bonus exercises depends on the matching of suitable exercises to the students.
 Most students found the exercises to match their weaknesses, independent of whether they received random or a tailored bonus exercises, probably caused by too few exercises available. 
 To validate the efficacy of our algorithm, we will increase the pool of bonus exercises which currently limits expressiveness of our findings.

High performing students were asking for more challenging bonus exercises.
We plan to extend our recommendation algorithm to detect high performing students and provide them with more challenging bonus exercises than struggling ones.
As discussed in~\cite{Teusner2017Aspects}, student performance classification is a tough research topic itself because there are very many variables to consider and assessments before the course starts are impractical.
We argue that we can use our knowledge model as a first step towards this topic.
We consider to incorporate typing patterns and writing speed of students into our knowledge model to determine their skill as discussed by Leinonen et al.~\cite{Leinonen2017Preventing}.
To further improve recommendation accuracy of exercises we will gather user feedback and working times to calibrate the item difficulty while the course is running.


\section{Conclusion}
We developed two concepts to support students in MOOCs and to improve their learning success: an automatic intervention system and a recommendation system for tailored bonus exercises.
For evaluation we conducted A/B tests on a live course with 5,839 active students.
We intervened students who were working on an exercise longer than 75\% of their peers.

While we were not able to reduce the dropout rates of students, we were able to show that interventions increased the amount of Requests For Comments (RFCs) by up to 66\% per student.
Students who received break interventions also showed an increased usage of RFCs even though we did not explicitly recommend it to them.
We could also show that students in the experimental groups reached out for help earlier than students in the control group.
Together with the surveys, which reassured us the helpfulness of RFCs for the students, we conclude that RFCs provide a benefit for students.
With RFCs most students were able to resolve their individual problems in less than five minutes.
Therefore, by increasing the amount of RFCs with interventions, we likely improved their learning experience.
We could show that interventions further caused students to do 25\% longer breaks on average which are likely to be beneficial for their learning process.
We further plan to improve intervention design and timings, as some students experienced the interventions as too disruptive or ignored them.
In addition to the intervention system, we also presented a content-based recommendation system which recommends bonus exercises to students at any given time.

In contradiction to our initial assumption, we were able to show a tendency that students perform better in their recommended exercises than those who got the same exercises by chance.
Perhaps students that spent more time with certain topics, even if they struggled, actually learned more than those who quickly went through without any problems.
With interventions and tailored bonus exercises we created two approaches to improve MOOCs by leveraging individualized learning.
This work thus offers another step towards personalized MOOCs by helping students to overcome individual content related problems.

%
%
%
%
%
\balance{}

\balance{}

\bibliographystyle{SIGCHI-Reference-Format}
\bibliography{citations}


\begin{thebibliography}{00}


\ifx \showCODEN    \undefined \def \showCODEN     #1{\unskip}     \fi
\ifx \showDOI      \undefined \def \showDOI       #1{{\tt DOI:}\penalty0{#1}\ }
  \fi
\ifx \showISBNx    \undefined \def \showISBNx     #1{\unskip}     \fi
\ifx \showISBNxiii \undefined \def \showISBNxiii  #1{\unskip}     \fi
\ifx \showISSN     \undefined \def \showISSN      #1{\unskip}     \fi
\ifx \showLCCN     \undefined \def \showLCCN      #1{\unskip}     \fi
\ifx \shownote     \undefined \def \shownote      #1{#1}          \fi
\ifx \showarticletitle \undefined \def \showarticletitle #1{#1}   \fi
\ifx \showURL      \undefined \def \showURL       #1{#1}          \fi

\bibitem{Agrawal2015Youedu}
{Akshay Agrawal}, {Jagadish Venkatraman}, {Shane Leonard}, {and} {Andreas
  Paepcke}.
\newblock \showarticletitle{{YouEDU: Addressing confusion in MOOC discussion
  forums by recommending instructional video clips}}. In {\em Proc. EDM '15}.
  Stanford InfoLab, 297--304.
\newblock


\bibitem{alevenLimitations2000}
{Vincent Aleven} {and} {Kenneth~R. Koedinger}. 2000.
\newblock \showarticletitle{Limitations of {Student} {Control}: {Do} {Students}
  {Know} when {They} {Need} {Help}?}. In {\em Intelligent {Tutoring}
  {Systems}}. Springer, Berlin, Heidelberg, 292--303.
\newblock
\showISBNx{978-3-540-67655-3 978-3-540-45108-2}


\bibitem{arigaBrief2011}
{Atsunori Ariga} {and} {Alejandro Lleras}. 2011.
\newblock \showarticletitle{Brief and rare mental ``breaks'' keep you focused:
  {Deactivation} and reactivation of task goals preempt vigilance decrements}.
\newblock {\em Cognition\/} {118}, 3 (March 2011), 439--443.
\newblock
\showISSN{0010-0277}


\bibitem{Bauman2014Recommending}
{Konstantin Bauman} {and} {Alexander Tuzhilin}. 2014.
\newblock \showarticletitle{{Recommending learning materials to students by
  identifying their knowledge gaps}}.
\newblock {\em CEUR Workshop\/}  {1247} (2014).
\newblock
\showISSN{16130073}


\bibitem{Carini2006Student}
{Robert~M. Carini}, {George~D. Kuh}, {and} {Stephen~P. Klein}. 2006.
\newblock \showarticletitle{{Student Engagement and Student Learning: Testing
  the Linkages}}.
\newblock {\em Research in Higher Ed.\/} {47}, 1 (2 2006).
\newblock
\showISSN{0361-0365}


\bibitem{Chandrasekaran2015Learning}
{Muthu~Kumar Chandrasekaran}, {Min-Yen Kan}, {Bernard C.~Y. Tan}, {and}
  {Kiruthika Ragupathi}. 2015.
\newblock \showarticletitle{{Learning Instructor Intervention from MOOC Forums:
  Early Results and Issues}}.
\newblock {\em Proc. LAK '17\/} (4 2015), 512--513.
\newblock


\bibitem{Chaturvedi2014Predicting}
{Snigdha Chaturvedi}, {Hal Daum}, {and} {Dan Goldwasser}. 2014.
\newblock \showarticletitle{{Predicting Instructor's Intervention in MOOC
  forums}}.
\newblock {\em Proceedings of the 52nd Association for Computational
  Linguistics\/} (2014), 1501--1511.
\newblock


\bibitem{He2015Identifying}
{Jiazhen He}, {James Bailey}, {Benjamin Rubinstein}, {and} {Rui Zhang}. 2015.
\newblock \showarticletitle{{Identifying At-Risk Students in Massive Open
  Online Courses}}.
\newblock {\em Conference on AI '15\/} (2015).
\newblock


\bibitem{Jadud2006Methods}
{Matthew~C Jadud}. 2006.
\newblock \showarticletitle{{Methods and tools for exploring novice compilation
  behaviour}}.
\newblock {\em Proc. Workshop on Computing Education Research\/} (2006),
  73--84.
\newblock


\bibitem{Jiang2014Predicting}
{Suhang Jiang}, {Adrienne~E Williams}, {Katerina Schenke}, {Mark Warschauer},
  {and} {Diane~O Dowd}. 2014.
\newblock \showarticletitle{{Predicting MOOC Performance with Week 1
  Behavior}}.
\newblock {\em Proc. EDM '14\/} (2014), 273--275.
\newblock


\bibitem{kizilcecDiverse2017}
{Ren{\'e}~F. Kizilcec} {and} {Christopher Brooks}. 2017.
\newblock \showarticletitle{Diverse {Big} {Data} and {Randomized} {Field}
  {Experiments} in {Massive} {Open} {Online} {Courses}}.
\newblock In {\em The {Handbook} of {Learning} {Analytics}} (1 ed.). Society
  for Learning Analytics Research (SoLAR), Alberta, Canada, 211--222.
\newblock
\showISBNx{978-0-9952408-0-3}


\bibitem{Kizilcec2017PNAS}
{Ren{\'e}~F. Kizilcec} {and} {Geoffrey~L. Cohen}. 2017.
\newblock \showarticletitle{Eight-minute self-regulation intervention raises
  educational attainment at scale in individualist but not collectivist
  cultures}.
\newblock {\em Proc. NAS\/} (2017).
\newblock
\showISSN{0027-8424}


\bibitem{Kizilcec2017affirmation}
{Ren{\'e}~F. Kizilcec}, {Glenn~M. Davis}, {and} {Geoffrey~L. Cohen}. 2017.
\newblock \showarticletitle{Towards Equal Opportunities in MOOCs: Affirmation
  Reduces Gender \& Social-Class Achievement Gaps in China}. In {\em Proc. L@S
  '17} {\em (L@S '17)}. ACM, New York, NY, USA, 121--130.
\newblock
\showISBNx{978-1-4503-4450-0}


\bibitem{KizilcecAttrition2015}
{Ren{\'e}~F. Kizilcec} {and} {Sherif Halawa}. 2015.
\newblock \showarticletitle{Attrition and Achievement Gaps in Online Learning}.
  In {\em Proc. L@S '15}. ACM, New York, NY, USA, 57--66.
\newblock
\showISBNx{978-1-4503-3411-2}


\bibitem{Kizilcec2015Motivation}
{Ren{\'{e}}~F. Kizilcec} {and} {Emily Schneider}. 2015.
\newblock \showarticletitle{{Motivation As a Lens to Understand Online
  Learners: Toward Data-Driven Design with the OLEI Scale}}.
\newblock {\em ACM Trans. Comput.-Hum. Interact.\/} {22}, 2 (3 2015), 1--24.
\newblock
\showISSN{1073-0516}


\bibitem{kizilcec2014encouraging}
{Ren{\'e}~F. Kizilcec}, {Emily Schneider}, {Geoffrey~L. Cohen}, {and}
  {Daniel~A. McFarland}. 2014.
\newblock \showarticletitle{Encouraging forum participation in online courses
  with collectivist, individualist and neutral motivational framings}.
\newblock {\em Proc. EMOOCS '14\/} (2014), 80--87.
\newblock


\bibitem{Kloft2014Predicting}
{Marius Kloft}, {Felix Stiehler}, {Zhilin Zheng}, {and} {Niels Pinkwart}. 2014.
\newblock \showarticletitle{{Predicting MOOC Dropout over Weeks Using Machine
  Learning Methods}}.
\newblock {\em Proc. EMNLP '14\/} (2014), 60--65.
\newblock
\showISSN{20737904}


\bibitem{Kulkarni2015}
{Chinmay~E. Kulkarni}, {Michael~S. Bernstein}, {and} {Scott~R. Klemmer}. 2015.
\newblock \showarticletitle{PeerStudio: Rapid Peer Feedback Emphasizes Revision
  and Improves Performance}. In {\em Proc L@S'15}. ACM, New York, NY, USA,
  75--84.
\newblock
\showISBNx{978-1-4503-3411-2}


\bibitem{Leinonen2017Preventing}
{Juho Leinonen}, {Petri Ihantola}, {and} {Arto Hellas}. 2017.
\newblock \showarticletitle{{Preventing Keystroke Based Identification in Open
  Data Sets}}. In {\em Proc. L@S '17}. ACM, 101--109.
\newblock


\bibitem{Michlik2010Exercises}
{Pavel Michlik} {and} {Maria Bielikova}. 2010.
\newblock \showarticletitle{{Exercises Recommending for Limited Time
  Learning}}.
\newblock {\em Procedia Computer Science\/} {1}, 2 (2010), 2821--2828.
\newblock
\showISSN{18770509}


\bibitem{Onah2014Dropout}
{D.F.O Onah}, {J Sinclair}, {and} {R. Boyatt}. 2014.
\newblock \showarticletitle{{Dropout Rates Of Massive Open Online Courses:
  Behavioural Patterns}}.
\newblock {\em Proc. EDULEARN14\/} (2014), 1--10.
\newblock
\showISSN{2340-1117}


\bibitem{Renz2016}
{Jan Renz}, {Daniel Hoffmann}, {Thomas Staubitz}, {and} {Christoph Meinel}.
  2016.
\newblock \showarticletitle{Using A/B Testing in MOOC Environments}. In {\em
  Proc. LAK '16}. ACM, 304--313.
\newblock
\showISBNx{978-1-4503-4190-5}


\bibitem{Segal2014Edurank}
{Avi Segal}, {Ziv Katzir}, {Kobi Gal}, {Guy Shani}, {and} {Bracha Shapira}.
  2014.
\newblock \showarticletitle{{EduRank: A Collaborative Filtering Approach to
  Personalization in E-learning}}.
\newblock {\em Proc. EDM '14\/} (2014), 68--75.
\newblock


\bibitem{Taylor2014Likely}
{Colin Taylor}, {Kalyan Veeramachaneni}, {and} {Una-May O'Reilly}. 2014.
\newblock \showarticletitle{{Likely to stop? Predicting Stopout in Massive Open
  Online Courses}}.
\newblock {\em CoRR\/} (8 2014), 25.
\newblock


\bibitem{Teusner2017Aspects}
{Ralf Teusner}, {Thomas Hille}, {and} {Christiane Hagedorn}. 2017a.
\newblock \showarticletitle{Aspects on Finding the Optimal Practical
  Programming Exercise for {MOOCs}}. In {\em Proc. FIE '17}.
\newblock


\bibitem{teusner2017InformedAction}
{Ralf Teusner}, {Kai-Adrian Rollmann}, {and} {Jan Renz}. 2017b.
\newblock \showarticletitle{Taking {Informed} {Action} on {Student} {Activity}
  in {MOOCs}}. In {\em Proc. L@S '17} {\em (L@{S} '17)}. ACM, 149--152.
\newblock
\showISBNx{978-1-4503-4450-0}


\bibitem{Simko2010Alef}
{Mari{\'a}n \v{S}imko}, {Michal Barla}, {and} {M{\'a}ria Bielikov{\'a}}. 2010.
\newblock \showarticletitle{{ALEF}: {A} {Framework} for {Adaptive}
  {Web}-{Based} {Learning} 2.0}.
\newblock In {\em Key {Competencies} in the {Knowledge} {Society}}. Springer,
  Berlin, Heidelberg, 367--378.
\newblock
\showISBNx{978-3-642-15377-8 978-3-642-15378-5}


\bibitem{Whitehill2015Beyond}
{Jacob Whitehill}, {Joseph Williams}, {Glenn Lopez}, {Cody Coleman}, {and}
  {Justin Reich}. 2015.
\newblock \showarticletitle{{Beyond Prediction: First Steps Toward Automatic
  Intervention in MOOC Student Stopout}}.
\newblock {\em Proc. EDM '15\/} (2015), 171--178.
\newblock


\bibitem{Yang2014Question}
{Diyi Yang}, {David Adamson}, {and} {Carolyn~Penstein Ros{\'e}}. 2014.
\newblock \showarticletitle{{Question Recommendation with Constraints for
  Massive Open Online Courses}}. {\em Proc. RecSys '14\/} (2014).
\newblock


\bibitem{Yang2013Turn}
{Diyi Yang}, {Tanmay Sinha}, {and} {David Adamson}. 2013.
\newblock \showarticletitle{{``Turn on, Tune in, Drop out'': Anticipating
  student dropouts in Massive Open Online Courses}}.
\newblock {\em Proc. NIPS Workshop on Data Driven Education\/} (2013), 1--8.
\newblock


\bibitem{Yeomans2017}
{Michael Yeomans} {and} {Justin Reich}. 2017.
\newblock \showarticletitle{Planning Prompts Increase and Forecast Course
  Completion in Massive Open Online Courses}. In {\em Proc. LAK '17}. ACM, New
  York, NY, USA, 464--473.
\newblock
\showISBNx{978-1-4503-4870-6}


\end{thebibliography}

\end{document}